\documentclass[titlepage,twoside,12pt]{elsart}
\usepackage{epsfig}
\usepackage[T1]{fontenc}
\usepackage{amsmath}
\usepackage{hyperref}
\begin{document}
\begin{frontmatter}


\title{Analytical description of Recurrence Plots of white noise and chaotic processes} 

\author{M. Thiel, M.C. Romano, J. Kurths}
\address{University of Potsdam, Germany}






\newcommand{\eps}{\varepsilon}
\newcommand{\D}{\displaystyle}
\renewcommand{\phi}{\varphi}

\nonfrenchspacing



\begin{abstract}
We present an analytical description of the distribution of diagonal lines in Recurrence Plots (RPs) for white noise and chaotic systems, and find that the latter one is linked to the correlation entropy. Further we identify two scaling regions in the distribution of diagonals for oscillatory chaotic systems that are hinged to two prediction horizons and to the geometry of the attractor. These scaling regions cannot be observed with the Grassberger-Procaccia algorithm. Finally, we propose methods to estimate dynamical invariants from RPs. 
\end{abstract}

\end{frontmatter}

\section{Introduction}The Recurrence Quantification Analysis (RQA) quantifies structures found in Recurrence Plots (RPs) to yield a deeper understanding of the underlying process of a given time series \cite{eckmann87,gao00}. Even though this method is widely applied \cite{norbert,norbert2,Webber,zbilut98a,zbilut98,holyst01,kurths94}, the scarce mathematical description is a main drawback. First steps in the direction of an analytical description were made by Faure et al. \cite{faure}, who gave analytical results for the cumulative distribution of diagonals $P^c_{\varepsilon}(l)$ in RPs in the case of chaotic maps and linked the slope of this distribution to the Kolmogorov-Sinai entropy. Gao and Cai \cite{gao00} related the distribution $P^c_{\varepsilon}(l)$ to the largest Lyapunov exponent and the information dimension.\\
In this paper we give an analytical expression for the distribution of diagonals in RP in the case of stochastic processes and extend the results of \cite{faure,gao00} to chaotic flows. Further we compare our approach with the Grassberger-Procaccia (G-P) algorithm \cite{Grassb} and show some advantages of the RP method estimating some invariants of the dynamics, such as the correlation entropy. One of the most remarkable differences between our approach and the G-P algorithm, is that we find two different scaling regions for chaotic flows, such as the R\"ossler system, instead of the single one obtained with the G-P algorithm. This new scaling region can be linked to the geometry of the attractor and defines another characteristic time scale of the system. Beyond we propose optimized measures for the identification of relevant structures in the RP.\\
The outline of this paper is as follows. In Sec.~\ref{2} we briefly introduce RPs. After considering in Sec.~\ref{3} the RPs of white noise, we proceed to general chaotic system (Sec.\ref{unknown}). Then, we exemplify our theoretical results for the R\"ossler system (Sec.\ref{Roesslersystem}) and present the two different scaling regions that characterize the system. Finally, we propose to estimate main characteristics of nonlinear systems from the RP which extends the importance of the RQA (Sec.~\ref{10}).

\section{Recurrence Plots and Recurrence Quantification Analysis\label{2}}
RPs were introduced to simply visualize the behavior of trajectories in phase space \cite{eckmann87}. Suppose we have a dynamical system represented by the trajectory $\{\vec x_i\}$ for $i=1,\ldots,N$ in a $d$-dimensional phase space. Then we compute the matrix
\begin{equation}\label{eq1}
\mathbf{R}_{i,\,j} = \Theta(\varepsilon-\left|\vec x_{i} - \vec x_{j}\right|), \quad \, 
i, j=1\dots N,
\end{equation}
where $\varepsilon$ is a predefined threshold and $\Theta(\cdot)$ is the Heaviside function \footnote{The norm used in Eq.~\ref{eq1} is in principle arbitrary. For theoretical reasons, that we will present later, it is preferable to use the maximum norm. However the numerical simulations of this paper are based on the Euclidian norm to make the results comparable with the literature. The theoretical results of this paper hold for both choices of the norm.}. The graphical representation of $\mathbf{R}_{i,\,j}$ called Recurrence Plot is yielded encoding the value one as ''black'' and zero as ''white'' point. A homogeneous plot with mainly single points may indicate a mainly stochastic system. Paling away from the main diagonal may indicate a drift i.e. non-stationarity of the time series. A main advantage of this method is that it allows to apply it to nonstationary data \cite{norbert}.\\
To quantify the structures that are found in RPs, the Recurrence Quantification Analysis (RQA) was proposed \cite{Webber}. There are different measures that can be considered in the RQA. One crucial point for these measures is the distribution of the lengths of the diagonal lines $P_{\varepsilon}(l)$  that are found in the plot. In the case of deterministic systems the diagonal lines mean that trajectories in the phase space are close to each other on time scales that correspond to the lengths of the diagonals. In the next sections we show that there is a relationship between $P_{\varepsilon}(l)$ and the correlation entropy. On the other hand we compute the distribution of diagonals for random processes to see that even in this case, there are some diagonals which can lead to pitfalls in the interpretation of the RQA because noise is inevitable in experimental systems. A more detailed discussion of this problem is given in \cite{thiel}.

\section{Results for white noise\label{3}}\label{whitenoise}

In this section we compute analytically the probability to find a black or recurrence point and the distribution of diagonals of length $l$ in the RP in the case of independent noise. The probability to find a recurrence point in the RP is given by 
\begin{equation}\label{pb}
P_b(\varepsilon)= \lim_{N \to \infty} \frac{1}{N^2}\sum_{i,j=1}^N\mathbf{R}_{i,\,j},
\end{equation}
and the probability to find a diagonal of at least length $l$ in the RP is defined as
\begin{equation}\label{p_cum}
P_{\varepsilon}^c(l)=\lim_{N \to \infty} \frac{1}{N^2}\sum_{i,j=1}^N \prod_{m=0}^{l-1}\mathbf{R}_{i+m,\,j+m}, 
\end{equation} 
where $c$ stands for cumulative. Note that $P_b(\varepsilon)=P_{\varepsilon}(1)$.\\
We consider a random variable $X$ with probability density $\rho(x)$. Suppose that $\{x_i\}$ for $i=1,\ldots,N$ is a realization of $X$ and we are interested in the distribution of the distances of each point to all other points of the time series. This can be done by computing the convolution of the density $\rho(\cdot)$  
\begin{equation}
R(x)=\rho(x)\ast\rho(x).
\end{equation}
$P_b(\varepsilon)$ is then gained by integrating $R(x)$ over $[-\varepsilon,\varepsilon]$
\begin{equation}\label{rate}
P_b(\varepsilon)=\int_{-\varepsilon}^{\varepsilon}R(x)dx=2\int_{0}^{\varepsilon}R(x)dx.
\end{equation}
Note that $P_b(\varepsilon)$ is invariant against shuffling of the data. 
For $[0,1]$ uniformly distributed noise, $R(x)$ is given by
\begin{equation}\label{distances}
R(x)=\begin{cases}1-\left|x\right| & \text{if  }\left|x\right|<1\\
0 & \text{else}\end{cases}
\end{equation}
and hence the probability $P_b(\varepsilon)$ for RPs and CRPs is given by
\begin{equation}\label{pb_noise}
P_b(\varepsilon)=2\varepsilon-\varepsilon^2+\Theta\left(\varepsilon-1\right)\left[1-2\varepsilon+\varepsilon^2\right]
\end{equation}
For Gaussian white noise one finds $P_b(\varepsilon)=\text{erf}\left(\frac{\varepsilon}{2\sigma}\right)$, where $\sigma$ is the standard deviation.
Now it is straightforward to compute $P_{\varepsilon}^c(l)$ in the in CRPs (in RPs only asymptotically). As the noise is independent, we obtain
\begin{equation}\label{decaypl}
P_{\varepsilon}^c(l)=P_b(\varepsilon)^l.
\end{equation}
The probability to find a recurrence point $P_b(\varepsilon)$ is in both RPs and CRPs independent of the preceeding point on the diagonal (except in the main diagonal). Eq.~(\ref{decaypl}) shows that the probability to find a line of length $l$ decreases exponentially with $l$.
For our example of uniformly distributed noise we get 
\begin{equation}\label{Fornoise}
P_{\varepsilon}^c(l)=\left(2\varepsilon-\varepsilon^2\right)^l.
\end{equation}
Note that in this case the exponential decay depends on $\varepsilon$. 

\section{Results for chaotic systems \label{unknown}}

 We present in this section an approach for chaotic systems. It is an extension of the results presented in \cite{faure} for chaotic maps and also covers general chaotic flows. To estimate the distribution of the diagonals in the RP, we start with the correlation integral \cite{Grassb2} 
\begin{equation}\label{corr_int}
C(\varepsilon)=\lim_{N\to\infty}\frac{1}{N^2}\times \{\text{number of pairs
 $(i,j)$ with}\left|\vec{x}_i-\vec{x}_j\right|<\varepsilon \}.
\end{equation}
Note that the definition of $P_b(\varepsilon)$ coincides with the definition of the correlation integral 
\begin{equation}
C(\varepsilon) =\lim_{N\to\infty}\frac{1}{N^2}\sum\limits_{i=1}^N \Theta\left(\left|\vec{x}_i-\vec{x}_j\right|-\varepsilon\right)\stackrel{\text{Eq.1}}{=}\lim_{N\to\infty}\frac{1}{N^2}\sum\limits_{i,j=1}^{N}\mathbf{R}_{i,j}=P_b(\varepsilon).
\end{equation}
 This fact allows to link the known results about the correlation integral and the structures in RPs.\\
We consider a trajectory $\vec x(t)$ in the basin of attraction of an attractor in the $d$-dimensional phase space and the state of the system is measured at time intervals $\tau$. Let $\{1,2,...,M(\varepsilon)\}$ be a partition of the attractor in boxes of size $\varepsilon$. Then $p(i_1,...,i_l)$ denotes the joint probability that $\vec x(t=\tau)$ is in the box $i_1$, $\vec x(t=2\tau)$ is in the box $i_2$, ..., and $\vec x(t=l\tau)$ is in the box $i_l$. The order-2 R\'enyi entropy \cite{renyi,Grassb0} is then defined as 
\begin{equation}
K_2=-\lim_{\tau\to 0}\lim_{\varepsilon \to 0}\lim_{l \to \infty}\frac{1}{l\tau}\ln \sum_{i_1,...,i_l} p^2(i_1,\dots,i_l).
\end{equation}
We can approximate $p(i_1,\ldots,i_l)$ by the probability $P_{t,l}(\vec x,\varepsilon)$ of finding a sequence of points in boxes of length $\varepsilon$ about $\vec x(t=\tau)$, $\vec x(t=2\tau)$, ..., $\vec x(t=l\tau)$. Assuming that the system is ergodic, which is always the case for chaotic systems as they are mixing, we obtain
\begin{equation}
\sum_{i_1,\ldots,i_l}p^2(i_1,\ldots,i_l)=
\frac{1}{N}\sum_{t=1}^N p_t(i_1,\ldots,i_l)\sim \frac{1}{N}\sum_{t=1}^N P_{t,l}(\vec x, \varepsilon),
\end{equation}
 where $p_t(i_1,\ldots,i_l)$ represents the probability of being in the box $i_1$ at time $t=\tau$, in the box $i_2$ at time $t=2\tau$, ... and in the box $i_l$ at time $t=l\tau$. Further we can express $P_{t,l}(\vec x,\varepsilon)$ by means of the recurrence matrix 
\begin{equation}
P_{t,l}(\vec x,\varepsilon)=\frac{1}{N}\sum_{s=1}^N \prod_{m=0}^{l-1}\Theta\left(\varepsilon-|\vec x_{t+m}-\vec x_{s+m}|\right)=\frac{1}{N}\sum_{s=1}^N \prod_{m=0}^{l-1}\mathbf{R}_{t+m,s+m}.
\end{equation}
Hence we obtain an estimator for the order-2 R\'enyi entropy by means of the RP
\begin{equation}
\hat{K_2}(\varepsilon,l)=\frac{1}{l\tau}\ln\underbrace{\left(\frac{1}{N^2}\sum_{t,s=1}^N \prod_{m=0}^{l-1} \mathbf{R}_{t+m,s+m}\right)}_{(\ast)}.
\end{equation}
Note that $(\ast)$ is the cumulative distribution of diagonal lines $P^c_{\varepsilon}(l)$ (Eq.~(\ref{p_cum})).
Therefore, if we represent $P^c_{\varepsilon}(l)$ in a logarithmic scale versus $l$ we should obtain a straight line with slope $-\hat{K}_2(\varepsilon)\tau$ for large $l$'s.\\
On the other hand, in the G-P algorithm the $l$-dimensional correlation integral is defined as
\begin{equation}\begin{split}
C_l(\varepsilon)=\lim_{N \to \infty}\frac{1}{N^2}\sum_{t,s=1}^N \Theta\left(\varepsilon - \left(\sum_{k=0}^{l-1}\left|\vec x_{i+k}- \vec x_{j+k}\right|^2\right)^{1/2}\right).
\end{split}\end{equation}
Grassberger and Procaccia \cite{Grassb4} state that due to the exponential divergence of the trajectories, requiring 
\begin{equation}\label{Naeherung}
\sum_{k=0}^{l-1}\left|\vec x_{i+k}-\vec x_{j+k}\right|^2 \le \varepsilon^2
\end{equation}
 is essentially equivalent to
\begin{equation}\label{bedin1}
\left|\vec x_{i+k}-\vec x_{j+k}\right| < \varepsilon\quad\text{for}\;\;k=1,\ldots,l
\end{equation}
which leads to the ansatz:
\begin{equation}\label{GPansatz}
C_l(\varepsilon)\sim \varepsilon^\nu \exp(-l\tau K_2).
\end{equation}
Further they make use of Takens embedding theorem \cite{Takens} and reconstruct the whole trajectory from $l$ measurements of any single coordinate. Hence they consider
\begin{equation}\begin{split}
\tilde C_l(\varepsilon)=\lim_{N \to \infty}\frac{1}{N^2}\sum_{t,s=1}^N \Theta\left(\varepsilon - \left(\sum_{k=0}^{l-1}\left|x_{i+k}- x_{j+k}\right|^2\right)^{1/2}\right)
\end{split}\end{equation}
and use the same ansatz Eq.~(\ref{GPansatz}) for $\tilde{C}_l(\varepsilon)$. Then, the G-P algorithm obtains an estimator of $K_2$ considering
\begin{equation}\label{eq19}
\tilde{K}_{2}(\varepsilon,l)=\frac{1}{\tau}\ln\frac{\tilde{C}_l(\varepsilon)}{\tilde{C}_{l+1}(\varepsilon)}.
\end{equation}
Due to the similarity of the RP approach to the G-P one, we state
\begin{equation}\label{Main}
P^c_{\varepsilon}(l)\simeq \sum_{i_1,...,i_l} p^2(i_1,...,i_l)\simeq \tilde C_l(\varepsilon) \sim \varepsilon^\nu \exp(-l\tau K_2).
\end{equation} 
The difference between both approaches is that in $P_{\varepsilon}^c(l)$ we further consider information about $l$ vectors, whereas in $\tilde C_l(\varepsilon)$ we have just information about $l$ coordinates. Besides this, in the RP approach $l$ is a length in the plot, whereas in the G-P algorithm it means the embedding dimension. As $K_2$ is defined for $l \to \infty$, the RP approach seems to be more appropriate than the G-P one, as it is always problematic to use very high embedding dimensions \cite{Ruelle}.\\
A further advantage of the RP method is that it does not make use of the approximation that Eq.~(\ref{Naeherung}) is essentially equivalent to Eq.~(\ref{bedin1}). The quantity that enters the RPs is directly linked to the conditions Eq.~(\ref{bedin1}) and hence uses one approximation less than the G-P method.\\
One open question for both methods is the determination of the scaling regions. It is somewhat subjective and makes a rigorous error estimation problematic. For the cases considered in this paper we have found that 10,000 data points assure reliable results for both methods. Even 5,000 data points allow for a reasonable estimation, whereas 3,000 data points or less yield very small scaling regions that are difficult to identify. However, the RP method is advantageous for the estimation of $K_2$ as the representation is more direct. The most important advantage is presented in the next section: RPs allow to detect a new scaling region in the R\"ossler attractor that cannot be observed with the G-P algorithm.

 \section{The R\"ossler System\label{Roesslersystem}}\label{roessler}
\begin{figure}[hhh]
\begin{center}
\includegraphics[width=0.65\textwidth]{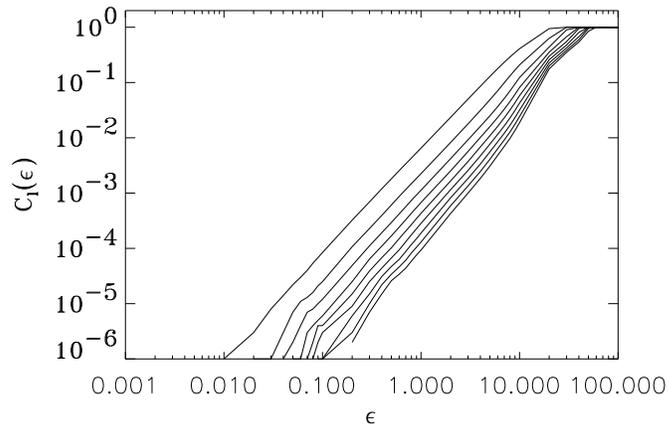}
\caption{\small G-P algorithm for the R\"ossler system. $l$ varies from $3$ (top) to $27$ (bottom) in steps of $3$.\label{GPRoessler}}
\end{center}
\end{figure}

\begin{figure}[hhh]
\begin{center}
\includegraphics[width=0.65\textwidth]{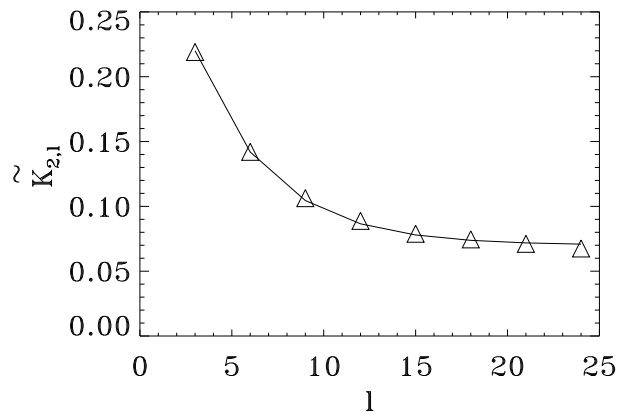}
\caption{\small Estimation of $K_{2,l}$ for the R\"ossler system with the G-P algorithm. The line is plotted to guide the eye.\label{K2GPRoessler}}
\end{center}
\end{figure}

\begin{figure}[hhh]
\begin{center}
\includegraphics[width=0.65\textwidth]{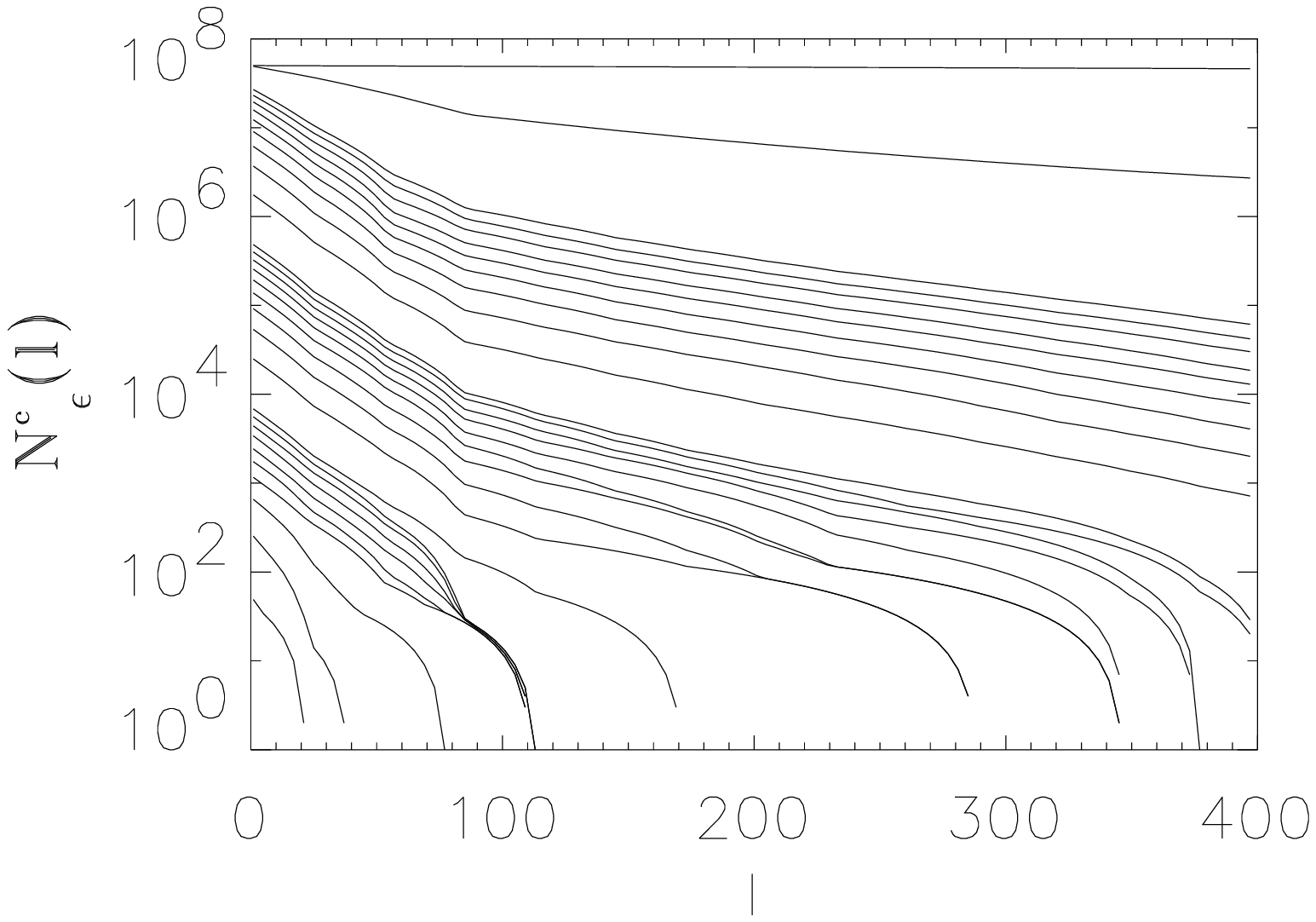}
\caption{\small RP method for the R\"ossler system. $\varepsilon$ varies logarithmically from $10^{-2}$ to $10.0$ (bottom to top)\label{RPRoessler}}
\end{center}
\end{figure}

\begin{figure}[hhh]
\begin{center}
\includegraphics[width=0.65\textwidth]{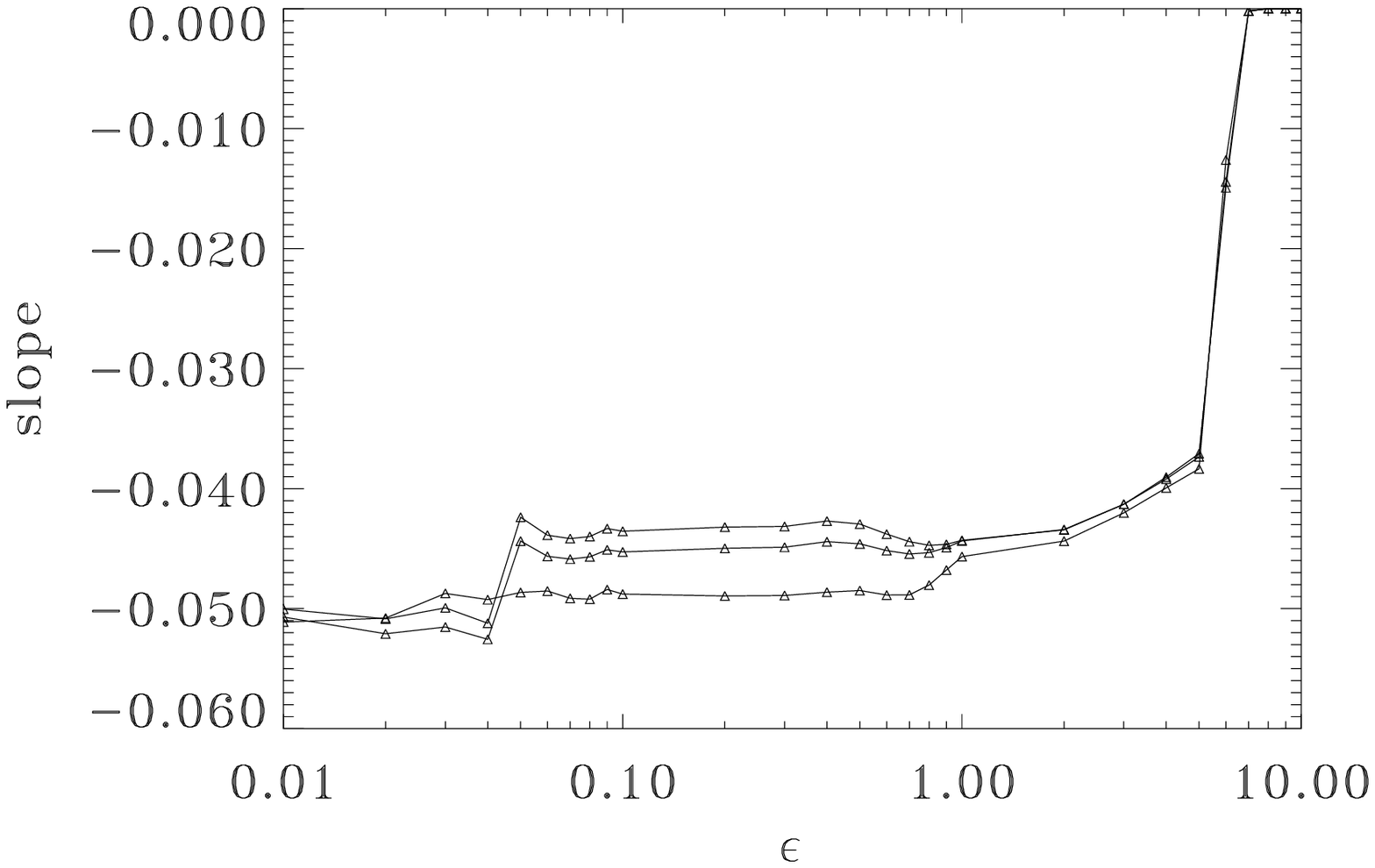}
\caption{\small RP method for the R\"ossler system: slope of the curves $N^c_{\varepsilon}(l)$ in the first region for three different choices of the scaling region in $l$.\label{Slope1Roessler}}
\end{center}
\end{figure}

\begin{figure}[hhh]
\begin{center}
\includegraphics[width=0.65\textwidth]{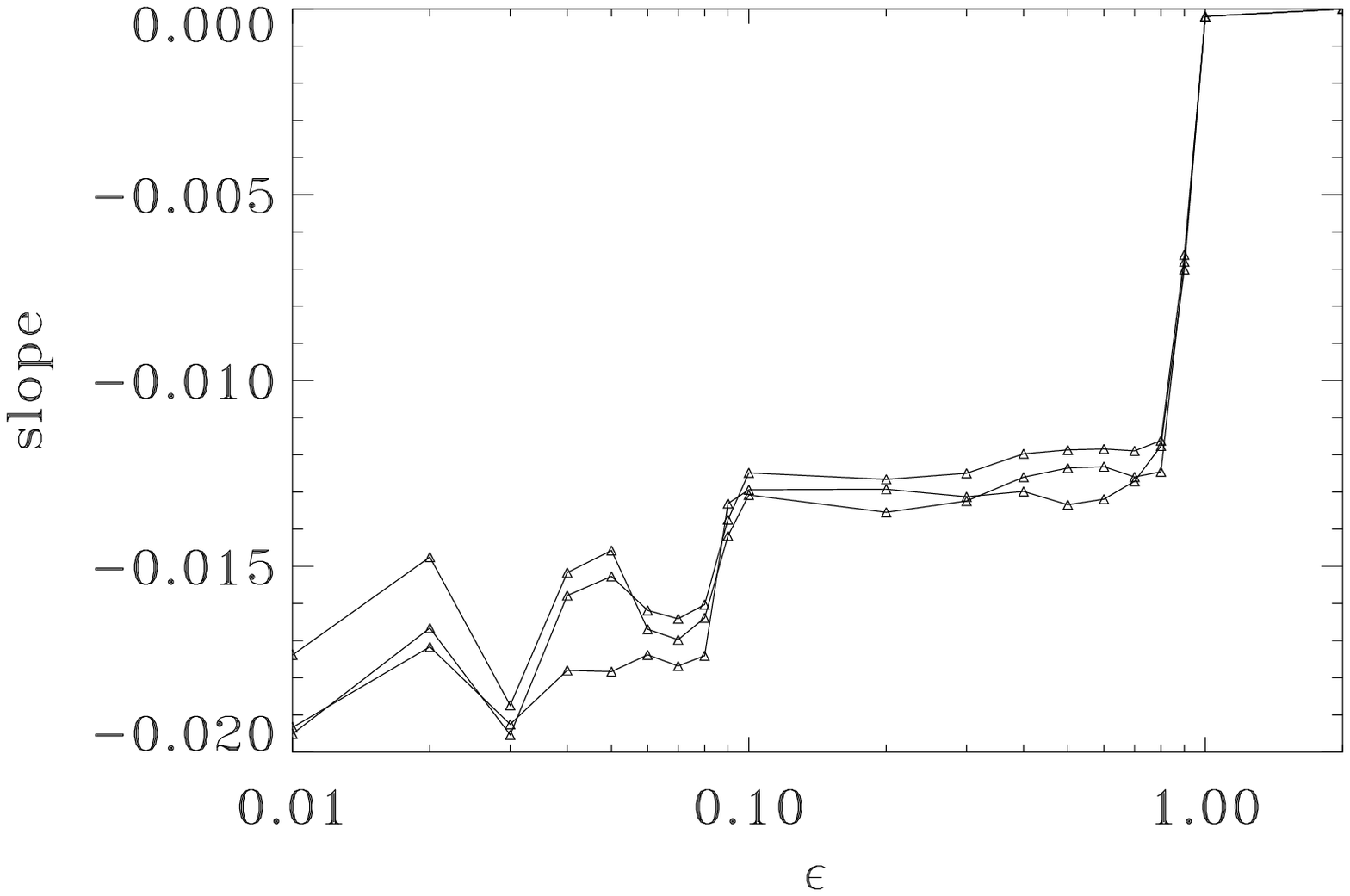}
\caption{\small RP method for the R\"ossler system: slope of the curves $N^c_{\varepsilon}(l)$ in the second region for three different choices of the scaling region in $l$\label{Slope2Roessler}}
\end{center}
\end{figure}

\begin{figure}[hhh]
\begin{center}
\includegraphics[width=0.65\textwidth]{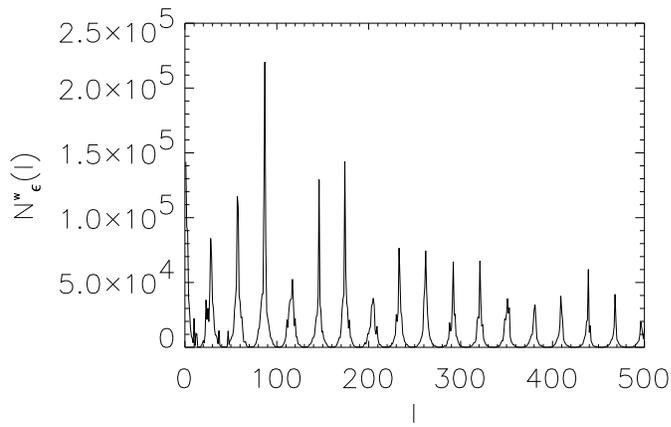}
\caption{\small Number of vertical white lines in the Recurrence Plot of the R\"ossler system with standard parameters, $\varepsilon=0.05$ and based on 60,000 data points.\label{RealRecurrence}}
\end{center}
\end{figure}

We analyze the R\"ossler system with standard parameters $a=b=0.2,c=5.7$ \cite{Roessler}. We generate 15,000 data points based on the Runge Kutta method of fourth order and neglect the first 5,000. The integration step is $h=0.01$ and the sampling rate is $20$.\\
First, we estimate $K_2$ by means of the G-P algorithm. Fig.~\ref{GPRoessler} shows the results for the correlation integral in dependence on $\varepsilon$.
There is one well-expressed scaling region for each embedding dimension $l$. Then we get from the vertical distances between the lines an estimate of $K_2$ (Fig.\ref{K2GPRoessler}), $\tilde{K}_{2}= 0.070\pm 0.003$.
Next, we calculate the cumulative distribution of the diagonal lines of the RP in dependence on the length of the lines $l$ (Fig.~\ref{RPRoessler}). 
For large $l$ and small $\varepsilon$ the scaling breaks down as there are not enough lines in the RP. The most remarkable fact in this figure is the existence of two well differentiated scaling regions. The first one is found for $1\le l \le 84$ and the second one for $l\ge 85$. The existence of two scaling regions is a new and striking point obtained from this analysis and is not observed with the G-P method. The estimate of $K^f_2$ from the slope of the first part of the lines is $K^f_2\approx 0.225\pm 0.03$ (Fig.~\ref{Slope1Roessler}) and the one from the second part is $K_2\approx 0.0675\pm 0.004$ (Fig.~\ref{Slope2Roessler}).
Hence, $K^f_2$ is between 3-4 times higher than $K_2$. As $K_2$ is defined for $l\to \infty$, the second slope yields the estimation of the entropy.\\
However, the first part of the curve is interesting too, as it is also independent of $\varepsilon$. The region $1\le l \le 84$ characterizes the short term dynamics of the system  up to three cycles around the fix point and corresponds in absolute units to a time of $t=16.8$, as we use a sampling rate of $\delta t=0.2$. These three cycles reflect a characteristic period of the system that we will call {\em recurrence period} $T_{\text{rec}}$. It is different from the dominant ``phase period'' $T_{\text{ph}}$, which is given by the dominant frequency of the power density spectrum. $T_{\text{rec}}$ however, is given by recurrences to the same state in phase space. 
Recurrences are represented in the plot by vertical (or horizontal, as the plot is symmetric) white lines. Such a white line occurs at the coordinates $i,j$ if
\begin{equation}\label{whiteline}
\mathbf{R}_{i,j+m}=\begin{cases} 1 \qquad\text{if}\quad m=-1\\
0 \qquad\text{for}\quad m\in\{0,\ldots,l-1\}\\
1  \qquad\text{if}\quad m=l.
\end{cases}
\end{equation}
The trajectory $\vec{x}_n$ for times $n=j-1,\ldots,j+l$ is compared to the point $\vec{x}_i$. Then the structure given by Eq.~\ref{whiteline} can be interpreted as follows. At time $n=j-1$ the trajectory falls within an $\varepsilon$-box of $\vec{x}_i$. Then for $n=j,\ldots,j+l-1$ it moves outside of the box, until at $n=j+l$ it recurs to the  $\varepsilon$-box of $\vec{x}_i$. Hence, the length of the white line is proportional to the time that the trajectory needs to recur close to $\vec{x}_i$.\\
In Fig.~\ref{RealRecurrence} we represent the distribution of white vertical lines in the RP. 
The period of about 28 points corresponds to $T_{\text{ph}}$. However, the highest peak is found at a lag of about 87 points (the second scaling region begins at $l=85$). This means that after this time most of the points recur close to their initial state. This time also defines the recurrence period $T_{\text{rec}}$. For the R\"ossler attractor with standard parameters we find $T_{\text{rec}}=3T_{\text{ph}}$.\\
For predictions on time scales below the recurrence period, $\tau^f=1/K^f_2$ is a better estimate of the prediction horizon than $\tau=1/K_2$. This interesting result means that the possibility to predict the next value within an $\varepsilon$-range is in the first part by a factor of more than $3$ times worse than it is in the second part, i.e. there exist two time scales that characterize the attractor. The first slope is greater than the second one because it is more difficult to predict the next step if we have only information about a piece the trajectory for less than the recurrence period. Once we have scanned the trajectory for more than $T_{\text{rec}}$, the predictability increases and the slope of $P^c_\varepsilon(l)$ in the logarithmic plot decreases. Hence the first slope, as well as the time scale at which the second slope begins, reveal important characteristics of the attractor.\\
To investigate how the length of the first scaling region depends on the form of the attractor, we have varied the parameter $c$ of the R\"ossler system with fixed $a=b=0.1$, so that different types of attractors appear \cite{alligood}. Especially we have studied the cases $c=9$, which yields $T_{\text{rec}}=2T_{\text{ph}}$, and  $c=30$, which gives $T_{\text{rec}}=4T_{\text{ph}}$. In both cases the length of the first scaling region corresponds as expected to $T_{\text{rec}}$.\\
On the other hand, the existence of the two scalings may be linked to the nonhyperbolic nature of the R\"ossler system for this attractor type, because the resulting two time scales have been also recently found by Anishchenko et al. based on a rather subtle method \cite{Anishchenko}.
This effect also is detectable in other oscillating nonhyperbolic systems like the Lorenz system and will be studied in more detail in a forthcoming paper.  

\section{Dynamical invariants for the RQA\label{10}}
\begin{figure}[hhh]
\begin{center}
\includegraphics[width=0.65\textwidth]{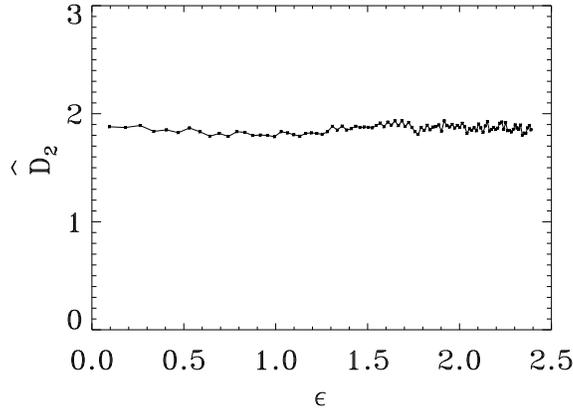}
\caption{\small Estimation of the correlation dimension $D_2$ for the R\"ossler attractor by the RP method. The parameters used for the R\"ossler system and the integration step are the same as in Sec.~\ref{roessler}.\label{CorrdimRP}}
\end{center}
\end{figure}

\begin{figure}[hhh]
\begin{center}
\includegraphics[width=0.65\textwidth]{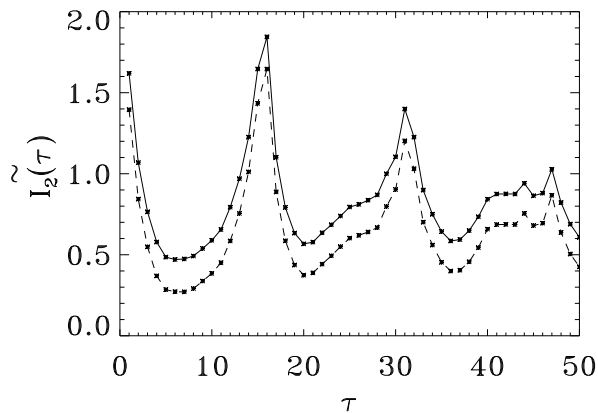}
\caption{\small Comparison of the estimators of the mutual information for the x-component of the R\"ossler system computed by the RP method (solid line) and the G-P algorithm (dashed line). The parameters used for the R\"ossler system and the integration step are the same as in Sec.~\ref{roessler}.\label{MI}}
\end{center}
\end{figure}

With regard to our theoretical findings in Sec.~\ref{unknown} we have to assess the quality of the possible results of the RQA.\\
The measures considered in the RQA \cite{Webber} are not invariants of the dynamical system, i.e. they usually change under coordinate transformations, and especially, they are in general modified by embedding \cite{thiel_chaos}. Hence, we propose new measures to quantify the structures in the RP, that are invariants of the dynamical system.\\
{\bf The first measure} we propose, is the slope of the cumulative distribution of the diagonals for large $l$. We have seen that it is (after dividing by $\tau$) an estimator of the R\'enyi entropy of second order $K_2$, which is a known invariant of the dynamics \cite{schuster}. On the other hand, we also can consider the slope of the distribution for small $l$'s, as this slope shows a clear scaling region, too. The inverse of these two quantities, is then related to the forecasting time at different horizons. Especially the transition point from the first to the second scaling region is an interesting characteristic of the system.\\
{\bf The second measure} we introduce, is the vertical distance between $P^c_{\varepsilon}(l)$ for different $\varepsilon$`s. From Eq.~(\ref{Main}) one can derive
\begin{equation}
\hat D_2(\varepsilon)=\ln\left(\frac{P^c_{\varepsilon}(l)}{P^c_{\varepsilon+\Delta\varepsilon}(l)}\right)\left(\ln\left(\frac{\varepsilon}{\varepsilon+\Delta\varepsilon}\right)\right)^{-1}.
\end{equation}
This is an estimator of the correlation dimension $D_2$ \cite{Grassb0}. The result for the R\"ossler system is represented in Fig.~\ref{CorrdimRP}. 
The mean value of $\hat D_2(\varepsilon)$ is in this case $1.86 \pm 0.04$. This result is in accordance with the estimation of $D_2$ by the G-P algorithm given in \cite{Raab}, where the value $1.81$ is obtained. With a modified G-P algorithm a value of $1.89$ was reported \cite{Raab}.\\
{\bf The third measure} we suggest, is an estimator of the generalized mutual information of order $2$,
\begin{equation}
I_2(\tau)=2H_2-H_2(\tau)
\end{equation}
where 
\begin{equation}
H_2=-\ln\sum\limits_{i}p_i^2,\qquad
H_2(\tau)=-\ln\sum\limits_{i}p_{i,j}^2(\tau)
\end{equation} 
are the generalized R\'enyi's second order entropy (also correlation entropy) and its corresponding joint second order entropy \cite{pompe}. This measure can be estimated using the G-P algorithm as follows \cite{kantz97}
\begin{equation}
\tilde I_2(\varepsilon,\tau)=\ln(C_2(\varepsilon,\tau))-2\ln(C_1(\varepsilon)).
\end{equation}
Instead, we can estimate $I_2(\tau)$ using the recurrence matrix. As discussed in the preceding sections, one can estimate $H_2$ as
\begin{equation}
\hat H_2 = -\ln \left[\frac{1}{N^2} \sum\limits_{i,j=1}^N \mathbf{R}_{i,j}\right]. 
\end{equation}
Analogously we can estimate the joint second order entropy by means of the recurrence matrix
\begin{equation}
\hat H_2(\tau) = -\ln \left[\frac{1}{N^2}\sum\limits_{i,j=1}^N \mathbf{R}_{i,j}\mathbf{R}_{i+\tau,j+\tau}\right]. 
\end{equation}   
We compare the estimation of $I_2(\tau)$ based on the G-P algorithm with the one obtained by the RP method in Fig.~\ref{MI}. 
We see, that the RP method yields systematically higher estimates of the mutual information, as in the case of the estimation of the correlation entropy. However, the structure of the curves is qualitatively the same (it is just shifted to higher values by about $0.2$). A more exhaustive inspection shows, that the difference is due to the use of the Euclidean norm. The estimate based on the RP method is almost independent of the norm, whereas the estimate based on the G-P algorithm clearly depends on the special choice. If the maximum norm is used (in G-P and RP) both curves coincide.\\
Note that the estimators for the invariants we propose are different from the ones of the G-P algorithm. Therefore, the obtained values are slightly different, too.\\
The three measures that we have proposed, are not only applicable for chaotic systems but also for stochastic ones as the invariants are equally defined for both kinds of systems.

\section{Conclusions}

In this paper we have presented an analytical expression for the the distribution of diagonals $P_{\varepsilon}(l)$ for stochastic systems and chaotic flows, extending the results presented in \cite{faure}. We have shown that $P_{\varepsilon}(l)$ is linked to the 2-order R\'enyi entropy rather than to the Lyapunov exponent. Further we have found in the logarithmic plot of $P_{\varepsilon}(l)$ two different scaling regions with respect to $\varepsilon$, that characterize the dynamical system and are also related to the geometry of the attractor. This is a new point that cannot be seen by the G-P algorithm and will be studied in more detail in a forthcoming paper. The first scaling region defines a new time horizon for the description of the system for short time scales. Beyond the RP method does not make use of high embedding dimensions, and the computational effort compared with the G-P algorithm is decreased. Therefore the RP method is rather advantageous than the G-P one for the analysis of rather small and/or noisy data sets. Besides this, we have proposed different measures for the RQA, like estimators of the second order R\'enyi entropy $K_2$, the correlation dimension $D_2$ and the mutual information, that are, in contrast to the usual ones, invariants of the dynamics \cite{thiel_chaos}.

\subsection*{Acknowledgments}
We thank Dieter Armbruster, Annette Witt, Udo Schwarz and Norbert Marwan for the fruitful discussions. The project was supported by the ''DFG-Schwerpunktprogramm 1114''.


\begin{appendix}

\end{appendix}



\end{document}